\documentclass[aps,prl,preprint,amsmath,amssymb,showpacs,superscriptaddress]{revtex4}

\usepackage{graphicx}
\usepackage{dcolumn}
\usepackage{bm}
\usepackage{longtable}
\usepackage{color}
\usepackage{CJK}

\begin{document}

\title{ Microscopic understanding for the simplicity of the quadrupole moments newly observed in Cd isotopes }

\author{P. W. Zhao}%
\affiliation{State Key Laboratory of Nuclear Physics and Technology, School of Physics, Peking University, Beijing 100871, China}
\author{S. Q. Zhang}%
\affiliation{State Key Laboratory of Nuclear Physics and Technology, School of Physics, Peking University, Beijing 100871, China}
\author{J. Meng}%
\affiliation{State Key Laboratory of Nuclear Physics and Technology, School of Physics, Peking University, Beijing 100871, China}
\affiliation{School of Physics and Nuclear Energy Engineering, Beihang University, Beijing 100191, China}
\affiliation{Department of Physics, University of Stellenbosch, Stellenbosch, South Africa}
\date{\today}
\begin{abstract}
The simplicity in the nuclear quadrupole moments reported recently in $^{107-129}$Cd, i.e., a linear increase of the ${11/2}^-$ quadrupole moments, is investigated microscopically with the covariant density functional theory. Using the newly developed functional PC-PK1, the quadrupole moments as well as their linear increase tendency with the neutron number are
excellently reproduced without any {\it ad hoc} parameters. The core polarization is found to be very important and contributes almost half of the quadrupole moments. The simplicity of the linear increase is revealed to be due to the pairing correlation which smears out the abrupt changes induced by the single-particle shell structure.
\end{abstract}


\maketitle


The electric and magnetic moments are fundamental observables in understanding matter structure ranging from physics to chemistry. In nuclear physics, due to the simple and well-known structure of the electromagnetic interaction, the study of the electric and magnetic moments and transition probabilities provide unique opportunity for detailed understanding of nuclear structure and has attracted the attention of nuclear
physicists since the early days~\cite{Bohr1969}.

The electric quadrupole moment is the most important multipole expansion of a nuclear electric moment, reflecting the extent to which the nuclear charge distribution deviates from sphericity and representing the most important collective excitations of the nucleus.
In fact, the electric quadrupole moment in deuteron~\cite{Kellogg1939Phys.Rev.318} provides the first evidence for the need of a tensor term in the nuclear interaction~\cite{Bethe1939Phys.Rev.1261}, which turns out to be essential in understanding the variation of the nuclear shell structure~\cite{Otsuka2005Phys.Rev.Lett.232502}.
Many hot topics in nuclear physics including the exotic deformation~\cite{Bark2010Phys.Rev.Lett.22501}, shape coexistence~\cite{Heyde2011Rev.Mod.Phys.1467}, and shell evolution~\cite{Chevrier2012Phys.Rev.Lett.162501} root in the study of the nuclear electric quadrupole moment.

The extreme single-particle shell model, which has played a major role in clarifying nuclear structure~\cite{Mayer1955}, is widely used to describe the nuclear quadrupole moments. In this model, the quadrupole moment of an odd-proton (neutron) nucleus with spin $I$ is determined by the unpaired proton (neutron) in the orbital with the total angular momentum $j=I$.
For nucleus with one particle/hole around the doubly magic nucleus, e.g., $^{208}\rm Pb$~\cite{Sagawa1988Phys.Lett.B15}, a good agreement with the experimental quadrupole moments
can be achieved with the introduction of the effective charges.
In such a way, the interaction between the valence nucleons and the core nucleons or the core polarization can be effectively taken into account.

For nucleus with $n$ valence particles (holes) in an orbital $j$ coupled to a total spin $I$, the shell model predicts
that the quadrupole moments of nuclear states described by such a pure configuration would follow a simple linear relation with respect to the number of valence protons~\cite{Mayer1955} or neutrons~\cite{Horie1955Phys.Rev.778}. In general, the corresponding formalism can be given in the seniority scheme as~\cite{Shalit1963}
\begin{equation}\label{SM}
  \langle j^n|\hat{Q}|j^n\rangle = \frac{2j+1-2n}{2j+1-2\nu}\langle j^\nu|\hat{Q}|j^\nu\rangle
\end{equation}
with $\nu$ being the number of the unpaired valence nucleons. Due to this very characteristic linear behavior with respect to the number of valence nucleons, the electric quadrupole moment is believed to provide an stringent test of the model.

Experimentally, however, there is still a large scarcity of the quadrupole moment data for such examinations. The only examples include the experimental quadrupole moments of the $i_{13/2}$ isomers of lead and mercury isotopes as discussed in Ref.~\cite{Neyens2003Rep.Prog.Phys.633}.

In a recent paper~\cite{Yordanov2013Phys.Rev.Lett.192501}, the quadrupole moments of neutron-rich isotopes of cadmium up to the $N=82$ shell closure have been investigated by high-resolution laser spectroscopy, and the linear signature is confirmed by a linear increase of the $11/2^-$ quadrupole moments and is found to act well even beyond the $h_{11/2}$ shell. It is the first to demonstrate that the linear behavior can persist even beyond a single shell, and the high experimental precision achieved provides a rigorous calibration for theories. As mentioned in Ref.~\cite{Wood2013Physics52}, the result reported by Yordanov \textit{et al.}~\cite{Yordanov2013Phys.Rev.Lett.192501} is a success for the shell model and the concept of pairing of neutrons and of protons to produce a simple and persistent structure in a long isotopic chain. The challenge to the theorists now is to explain this simplicity in a microscopic level.

The density functional theory (DFT) plays a very important role in describing the many-body problems in a microscopic way. It has achieved great successes in all quantum mechanical many-body systems, e.g., in Coulombic systems. In nuclear physics, with spin and isospin degrees of freedom, the situation is much more complicated due to the strong nucleon-nucleon forces.
As Lorentz invariance is one of the underlying symmetries for the quantum
chromodynamics of the strong interaction, the covariant density
functionals~\cite{Ring1996Prog.Part.Nucl.Phys.193,Vretenar2005Phys.Rep.101,Meng2006Prog.Part.Nucl.Phys.470, Meng2013FrontiersofPhysics55} are of particular interest in nuclear physics. This symmetry not only provides us a consistent treatment of the spin degrees of freedom, but also puts stringent restrictions on the number of parameters in the corresponding functionals without reducing the quality
of the agreement with experimental data~\cite{Ring2012PhysicaScripta14035}.

In this Letter, the quadrupole moments of the Cd isotopes together with their linear simplicity shown in the most up-to-date data will be investigated in the framework of covariant density functional theory (CDFT) in a self-consistent and microscopic way. The major advantages of the present framework include: (1) the core polarization effects are fully taken into account and there is no need to introduce a factitious effective charge; (2) the pairing correlation is treated self-consistently by the corresponding pairing functional; (3) the density functional is universal for all nuclei throughout the periodic chart, and the present investigation is expected to be reliable and to have predictive power; (4) a microscopic picture of the nuclear quadrupole moments can be provided in terms of
intrinsic shapes and single-particle shells self-consistently.


Covariant density functional theory starts from an effective Lagrangian and the corresponding Kohn-Sham equations have the form of Dirac equation:
\begin{equation}\label{Diracequation}
   [\bm{\alpha}\cdot\bm{p}+\beta(m+S)+V]\psi_k=\epsilon_k\psi_k,
 \end{equation}
 with effective potentials $S(\bm{r})$ and $V(\bm{r})$ derived from this Lagrangian, which are connected in a self-consistent way to the densities, for details see Refs.~\cite{Ring1996Prog.Part.Nucl.Phys.193,Vretenar2005Phys.Rep.101,Meng2006Prog.Part.Nucl.Phys.470,Meng2013FrontiersofPhysics55}. The iterative solution of this equation yields the single-particle energies and wave functions, binding energies, quadrupole moments, etc. One should be noted that the calculations here are carried out in the intrinsic frame, and the measured quadrupole moment $Q$ for an axially symmetric nucleus is related to the intrinsic quadrupole moment $Q_0$ through the relation,
\begin{equation}\label{Eq:QM}
  Q = \frac{3K^2-I(I+1)}{(I+1)(2I+3)}Q_0,
\end{equation}
where $K$ is the projection of the total spin $I$ onto the symmetry axis of the deformed nucleus.


The Dirac equation (\ref{Diracequation}) is solved on the basis of an axially symmetric harmonic oscillator potential~\cite{Ring1997Comput.Phys.Commun.77} with 14 major shells. The point-coupling functional PC-PK1~\cite{Zhao2010Phys.Rev.C54319} is used for the Lagrangian
without any additional parameters, and the pairing correlations are taken into account by the Bardeen-Cooper-Schrieffer (BCS) method with a zero-range $\delta$ force.
By carrying out calculations with the triaxial degree of freedom, it is found that the triaxiality for the present Cd isotopes is negligible. Therefore, the effective potentials $S$ and $V$ are considered to be axially deformed.

Since we are focusing on the $11/2^-$ quadrupole moments of the odd-$A$ Cd isotopes observed in Ref.~\cite{Yordanov2013Phys.Rev.Lett.192501}, the last unpaired neutron will block its occupied level in the BCS calculations, i.e., the Pauli principle prevents this level from the scattering process of nucleon pairs by the pairing correlations. In practical calculations, the single-particle orbital $h_{11/2}$ with the third projection of the total angular momentum $j_z=11/2$ is always blocked in order to obtain the nuclear states with $K=11/2$.


\begin{figure}[htbp]
\includegraphics[width=7cm]{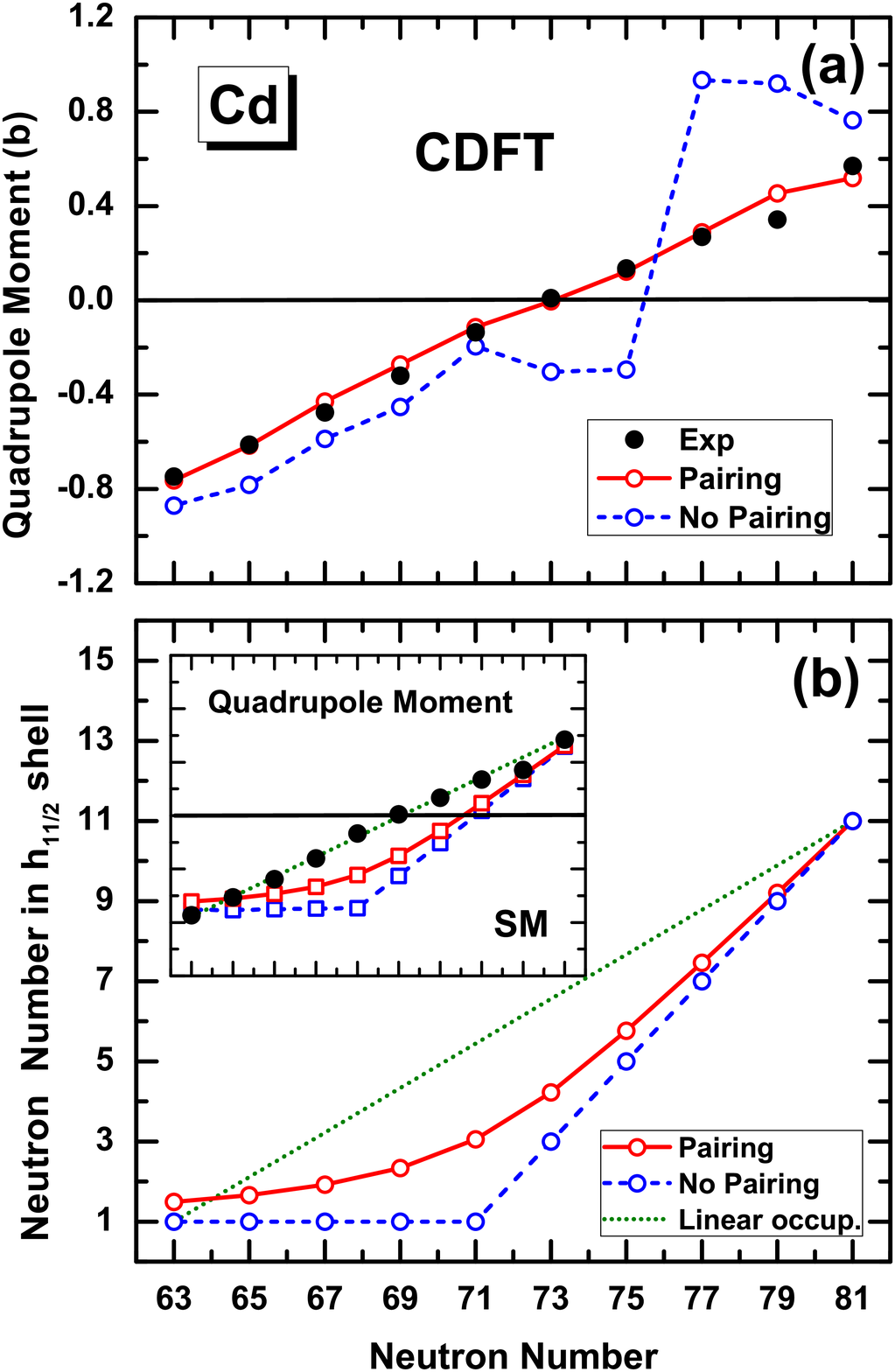}
\caption{(color online) Calculated quadrupole moments (a) and the occupation neutrons in the $h_{11/2}$ shell (b) in the covariant density functional theory (CDFT) with (full lines) and without (dashed lines) pairing correlation for Cd isotopes. The data (solid dots) are taken from Ref.~\cite{Yordanov2013Phys.Rev.Lett.192501}. The dotted line in panel (b) represents a linear occupation assumption proposed in the same reference~\cite{Yordanov2013Phys.Rev.Lett.192501}. Inset: Comparison between the observed quadrupole moments~\cite{Yordanov2013Phys.Rev.Lett.192501} and the calculated ones with Eq.~(\ref{SM}), in which the number of neutrons $n$ is either the CDFT results with (full line) and without (dashed line) pairing correlation or from the linear occupation assumption~\cite{Yordanov2013Phys.Rev.Lett.192501}.}
\label{fig1}
\end{figure}

In the upper panel of Fig.~\ref{fig1}, the calculated quadrupole moments in CDFT with and without pairing correlation are shown in comparison with the data~\cite{Yordanov2013Phys.Rev.Lett.192501}, respectively. One can see that, when the pairing correlations are taken into account, the experimental quadrupole moments together with the corresponding linear increase tendency are excellently reproduced in a microscopic and self-consistent way. Without pairing, however, one can clearly find two distinct features: for nuclei with $N\le71$, the calculated quadrupole moments are slightly smaller than the observed ones but still holds the linear increase tendency; for nuclei with $N\ge71$, the calculated results deviate not only from the data but also from the linear tendency. This indicates that the mechanism for the linear increase of the quadrupole moments may be different for nuclei with their neutron number smaller and larger than 71.

In the lower panel of Fig.~\ref{fig1}, similar features can also be seen for the neutron occupation number in the $h_{11/2}$ shell. Without pairing, there are always one neutron sitting in the $h_{11/2}$ shell for nuclei with $N\le71$, and this number starts to increase linearly from one for $^{119}$Cd to eleven for $^{129}$Cd. As a result, in the case without pairing, the increase of the neutrons in the $h_{11/2}$ shell shows a clear inflexion point at the nucleus $^{119}$Cd, which is consistent with the varying features of the quadrupole moments as shown in Fig.~\ref{fig1}(a). Such an abrupt inflexion point can be smoothed admirably by the pairing effects. Obviously, the neutron occupation number in $h_{11/2}$ shell with the pairing correlation does not follow the linear occupation assumption (dotted line in Fig.~\ref{fig1}(b)) proposed in Ref.~\cite{Yordanov2013Phys.Rev.Lett.192501}.

In the inset of Fig.~\ref{fig1}(b), adopting the linear occupation assumption,
the experimental quadrupole moments together with the corresponding linear increase tendency
can also be reproduced by Eq.~(\ref{SM}).
In the calculation, the single-particle quadrupole moment $\langle j^\nu|\hat{Q}|j^\nu\rangle|_{\nu = 1}$ of the $h_{11/2}$ shell in Eq.~(\ref{SM}) is estimated by $-e_{f}\langle r^2\rangle (2j-1)/(2j+2)$ with the effective charge $e_f = 2.5 e$ and
$ \langle r^2\rangle $ being calculated in CDFT, and it
is found to be nearly constant for all the Cd isotopes.
This explanation, however, is based on the linear occupation assumption shown as a dotted line in Fig.~\ref{fig1}(b). By replacing this fictitious occupation with the microscopic and self-consistent
occupation obtained from CDFT, one can easily find that the calculated results fail to reproduce the data, as shown in the inset of Fig.~\ref{fig1}(b), no matter whether the pairing correlation is included or not.
It should be mentioned that the calculated quadrupole moments here, similarly as in Ref.~\cite{Yordanov2013Phys.Rev.Lett.192501}, has been scaled with an small offset term $Q_{\rm const}$  around 150 mb.

\begin{figure*}[htbp]
\includegraphics[width=11cm]{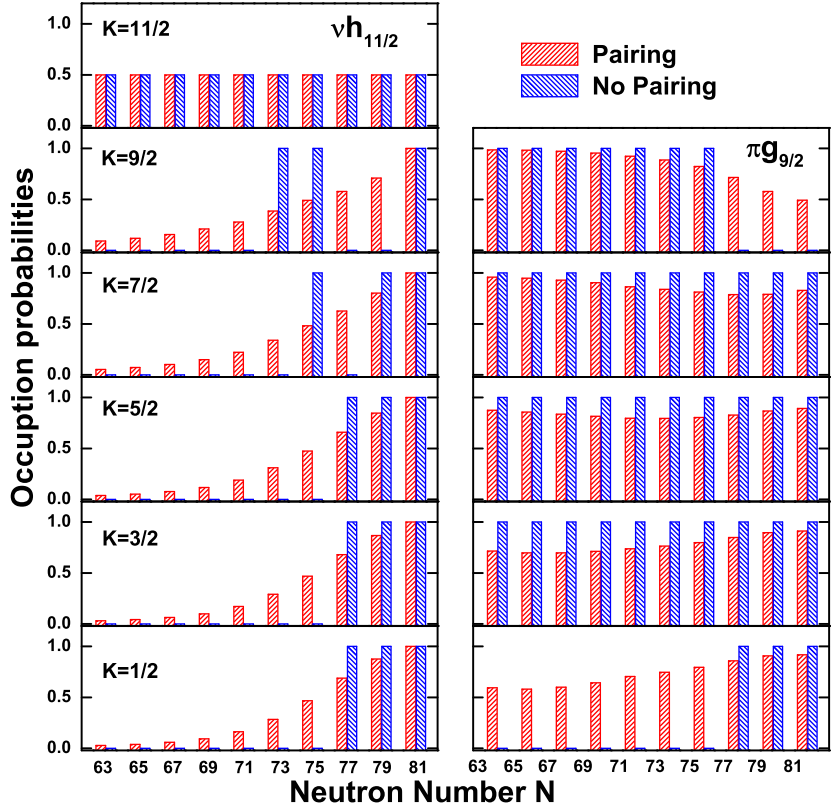}
\caption{(color online) Occupation probabilities for each single-particle orbital with $K$, the third component of the angular momentum in the neutron $\nu h_{11/2}$ shell (left panels) and the proton $\pi g_{9/2}$ shell (right panels), obtained from the CDFT calculations with and without pairing correlation.}
\label{fig2}
\end{figure*}

In the microscopic CDFT calculation, the quadrupole moment comes from all the protons in a nucleus. In order to understand the microscopic origin for the observed linear behavior of the quadrupole moments, we show in the left and right panels of Fig.~\ref{fig2} respectively the occupation probabilities for each single-particle orbital with $K$ the third component of the angular momentum in the neutron $\nu h_{11/2}$ shell and the proton $\pi g_{9/2}$ shell. Note that the degeneracy of each orbital here is two since a orbital with $K$ has the same energy as that with $-K$.

Without pairing correlation, since the neutron orbital $K=11/2$ is blocked, there are always one neutron in this orbital.
The occupation of neutron in other orbitals can be classified into two distinct cases.
For nuclei with $N\le71$, the increasing neutrons occupy other low $j$ shells instead of $h_{11/2}$ shell.
Due to the occupation of the high $K=11/2$ orbital, these nuclei have their neutron density distributions of oblate shapes.
Moreover, because there is a closure of the positive parity subshell at $N=70$, the neutron density distributions are driven toward a spherical shape with the increase of the neutron number.
Because of the attractive proton and neutron interaction, the proton density distributions evolve in a similar manner as the neutron ones with the increasing neutron number, and thus the quadrupole moments will approach zero smoothly.

For nuclei with $N\ge71$, it shows that the neutrons favor high $K$ orbitals for nuclei with $N\le75$, while favor low $K$ orbitals for nuclei with $N\ge77$. This indicates that with the increasing neutron number, the neutron density distribution changes from an oblate shape to a prolate shape near $N=76$. Similar evolution appears for the protons as well. It leads to an abrupt change of the single-proton level structure, and accordingly an abrupt jump of the quadrupole moments near $N=76$ as shown in Fig.~\ref{fig1}(a).

With pairing correlation, however, the occupation probabilities for all orbitals vary smoothly. It shows that the proton occupation of the low $K$ orbitals increases smoothly, while that of the high $K$ ones decline. This is connected with the nuclear shape evolution from an oblate to a prolate. As a result, one can conclude that the pairing effects can smear out all the abrupt changes in single-particle structure existing in the results without pairing, and thus leads to a smooth and gradual shape evolution from an oblate to a prolate. In such a way, the quadrupole moments increase in an almost linear way with the increasing neutron number and cross zero in the middle.

\begin{figure}[htbp]
\includegraphics[width=8cm]{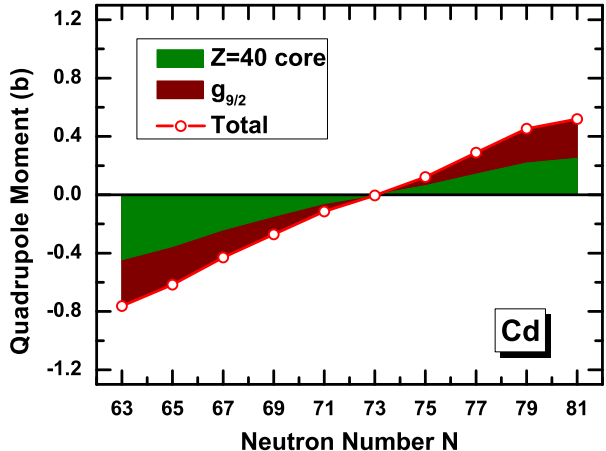}
\caption{(color online) Composition of the total quadrupole moments as a function of the neutron number for the Cd isotopes. It is shown that the $Z=40$ core contributes almost the same as the $g_{9/2}$ valence protons to the total quadrupole moments. }
\label{fig3}
\end{figure}

This is, however, not the full story. In order to investigate the importance of the core polarization effects, which is fully taken into account in the present work, one can extract the ``corelike'' quadrupole moments by excluding the contributions of valence nucleons, as shown in Fig.~\ref{fig3}, from to the total quadrupole moments. Since a full model space is taken in our calculations, we do not need to introduce the effective charges. Instead, the free charge is used in the present calculation, and thus the quadrupole moments are completely from the protons. Since $Z=40$ is a typical subshell closure, it is straightforward to regard the 48 protons in Cd isotopes as a $Z=40$ core coupled to 8 valence protons in $g_{9/2}$ shell.

In Fig.~\ref{fig3}, the composition of the total quadrupole moments are shown as a function of the neutron number for the Cd isotopes. It is found that the core contributes almost the same as the valence protons to the total quadrupole moments. Moreover, both the quadrupole moments from the core and valence proton follow a linear increase behavior. This demonstrates that the core is strongly polarized and its coupling with the valence protons is very remarkable. Therefore, it is of importance to emphasize that polarization effects play a very important role in the self-consistent microscopic description of nuclear quadrupole moments.

In summary, for the first time, the simplicity in the nuclear quadrupole moments has been explained in a fully self-consistent and microscopic way with the covariant density functional theory. The newly observed quadrupole moments data for $^{107-129}$Cd together with their linear increase tendency with the neutron number are excellently reproduced without any {\it ad hoc} parameters.
Instead of the linear neutron occupation mechanism in $h_{11/2}$ shell, the simplicity of the linear increase is revealed to be due to the pairing correlation which smears out the abrupt changes induced by the single-particle shell structure, and thus leads to a smooth shape evolution. Furthermore, it is found that the core is strongly coupled with the valence nucleons, and thus the core polarization effects are turned out to be essential which contribute almost half of the total quadrupole moments.

The authors are grateful to Q. B. Chen, Z. H. Zhang, and H. Z. Liang for helpful discussions. This work was partly supported by the Major State 973 Program 2013CB834400, the NSFC (Grants No. 11175002, No. 11105005, No. 11335002, No. 11345004, and No. 11375015), the Research Fund for the Doctoral Program of Higher Education under Grant No. 20110001110087, and the China Postdoctoral Science Foundation Grants No. 2012M520101 and No. 2013T60022.


\begin{thebibliography}{21}
\expandafter\ifx\csname natexlab\endcsname\relax\def\natexlab#1{#1}\fi
\expandafter\ifx\csname bibnamefont\endcsname\relax
  \def\bibnamefont#1{#1}\fi
\expandafter\ifx\csname bibfnamefont\endcsname\relax
  \def\bibfnamefont#1{#1}\fi
\expandafter\ifx\csname citenamefont\endcsname\relax
  \def\citenamefont#1{#1}\fi
\expandafter\ifx\csname url\endcsname\relax
  \def\url#1{\texttt{#1}}\fi
\expandafter\ifx\csname urlprefix\endcsname\relax\def\urlprefix{URL }\fi
\providecommand{\bibinfo}[2]{#2}
\providecommand{\eprint}[2][]{\url{#2}}

\bibitem[{\citenamefont{Bohr and Mottelson}(1969)}]{Bohr1969}
\bibinfo{author}{\bibfnamefont{A.}~\bibnamefont{Bohr}} \bibnamefont{and}
  \bibinfo{author}{\bibfnamefont{B.~R.} \bibnamefont{Mottelson}},
  \emph{\bibinfo{title}{Nuclear Structure}}, vol.~\bibinfo{volume}{I}
  (\bibinfo{publisher}{W. A. Benjamin, Inc.}, \bibinfo{year}{1969}).

\bibitem[{\citenamefont{Kellogg et~al.}(1939)\citenamefont{Kellogg, Rabi,
  Ramsey, and Zacharias}}]{Kellogg1939Phys.Rev.318}
\bibinfo{author}{\bibfnamefont{J.~M.~B.} \bibnamefont{Kellogg}},
  \bibinfo{author}{\bibfnamefont{I.~I.} \bibnamefont{Rabi}},
  \bibinfo{author}{\bibfnamefont{N.~F.} \bibnamefont{Ramsey}},
  \bibnamefont{and} \bibinfo{author}{\bibfnamefont{J.~R.}
  \bibnamefont{Zacharias}}, \bibinfo{journal}{Phys. Rev.}
  \textbf{\bibinfo{volume}{55}}, \bibinfo{pages}{318} (\bibinfo{year}{1939}).

\bibitem[{\citenamefont{Bethe}(1939)}]{Bethe1939Phys.Rev.1261}
\bibinfo{author}{\bibfnamefont{H.~A.} \bibnamefont{Bethe}},
  \bibinfo{journal}{Phys. Rev.} \textbf{\bibinfo{volume}{55}},
  \bibinfo{pages}{1261} (\bibinfo{year}{1939}).

\bibitem[{\citenamefont{Otsuka et~al.}(2005)\citenamefont{Otsuka, Suzuki,
  Fujimoto, Grawe, and Akaishi}}]{Otsuka2005Phys.Rev.Lett.232502}
\bibinfo{author}{\bibfnamefont{T.}~\bibnamefont{Otsuka}},
  \bibinfo{author}{\bibfnamefont{T.}~\bibnamefont{Suzuki}},
  \bibinfo{author}{\bibfnamefont{R.}~\bibnamefont{Fujimoto}},
  \bibinfo{author}{\bibfnamefont{H.}~\bibnamefont{Grawe}}, \bibnamefont{and}
  \bibinfo{author}{\bibfnamefont{Y.}~\bibnamefont{Akaishi}},
  \bibinfo{journal}{Phys. Rev. Lett.} \textbf{\bibinfo{volume}{95}},
  \bibinfo{pages}{232502} (\bibinfo{year}{2005}).

\bibitem[{\citenamefont{Bark et~al.}(2010)\citenamefont{Bark, Sharpey-Schafer,
  Maliage, Madiba, Komati, Lawrie, Lawrie, Lindsay, Maine, Mullins
  et~al.}}]{Bark2010Phys.Rev.Lett.22501}
\bibinfo{author}{\bibfnamefont{R.~A.} \bibnamefont{Bark}},
  \bibinfo{author}{\bibfnamefont{J.~F.} \bibnamefont{Sharpey-Schafer}},
  \bibinfo{author}{\bibfnamefont{S.~M.} \bibnamefont{Maliage}},
  \bibinfo{author}{\bibfnamefont{T.~E.} \bibnamefont{Madiba}},
  \bibinfo{author}{\bibfnamefont{F.~S.} \bibnamefont{Komati}},
  \bibinfo{author}{\bibfnamefont{E.~A.} \bibnamefont{Lawrie}},
  \bibinfo{author}{\bibfnamefont{J.~J.} \bibnamefont{Lawrie}},
  \bibinfo{author}{\bibfnamefont{R.}~\bibnamefont{Lindsay}},
  \bibinfo{author}{\bibfnamefont{P.}~\bibnamefont{Maine}},
  \bibinfo{author}{\bibfnamefont{S.~M.} \bibnamefont{Mullins}},
  \bibnamefont{et~al.}, \bibinfo{journal}{Phys. Rev. Lett.}
  \textbf{\bibinfo{volume}{104}}, \bibinfo{pages}{022501}
  (\bibinfo{year}{2010}).

\bibitem[{\citenamefont{Heyde and Wood}(2011)}]{Heyde2011Rev.Mod.Phys.1467}
\bibinfo{author}{\bibfnamefont{K.}~\bibnamefont{Heyde}} \bibnamefont{and}
  \bibinfo{author}{\bibfnamefont{J.~L.} \bibnamefont{Wood}},
  \bibinfo{journal}{Rev. Mod. Phys.} \textbf{\bibinfo{volume}{83}},
  \bibinfo{pages}{1467} (\bibinfo{year}{2011}).

\bibitem[{\citenamefont{Chevrier et~al.}(2012)\citenamefont{Chevrier, Daugas,
  Gaudefroy, Ichikawa, Ueno, Hass, Haas, Cottenier, Aoi, Asahi
  et~al.}}]{Chevrier2012Phys.Rev.Lett.162501}
\bibinfo{author}{\bibfnamefont{R.}~\bibnamefont{Chevrier}},
  \bibinfo{author}{\bibfnamefont{J.~M.} \bibnamefont{Daugas}},
  \bibinfo{author}{\bibfnamefont{L.}~\bibnamefont{Gaudefroy}},
  \bibinfo{author}{\bibfnamefont{Y.}~\bibnamefont{Ichikawa}},
  \bibinfo{author}{\bibfnamefont{H.}~\bibnamefont{Ueno}},
  \bibinfo{author}{\bibfnamefont{M.}~\bibnamefont{Hass}},
  \bibinfo{author}{\bibfnamefont{H.}~\bibnamefont{Haas}},
  \bibinfo{author}{\bibfnamefont{S.}~\bibnamefont{Cottenier}},
  \bibinfo{author}{\bibfnamefont{N.}~\bibnamefont{Aoi}},
  \bibinfo{author}{\bibfnamefont{K.}~\bibnamefont{Asahi}},
  \bibnamefont{et~al.}, \bibinfo{journal}{Phys. Rev. Lett.}
  \textbf{\bibinfo{volume}{108}}, \bibinfo{pages}{162501}
  (\bibinfo{year}{2012}).

\bibitem[{\citenamefont{Mayer and Jensen}(1955)}]{Mayer1955}
\bibinfo{author}{\bibfnamefont{M.~G.} \bibnamefont{Mayer}} \bibnamefont{and}
  \bibinfo{author}{\bibfnamefont{J.}~\bibnamefont{Jensen}},
  \emph{\bibinfo{title}{Elementary Theory of Nuclear Shell Structure}}
  (\bibinfo{publisher}{John Wiley \& Sons, Inc. New York},
  \bibinfo{year}{1955}).

\bibitem[{\citenamefont{Sagawa and Arima}(1988)}]{Sagawa1988Phys.Lett.B15}
\bibinfo{author}{\bibfnamefont{H.}~\bibnamefont{Sagawa}} \bibnamefont{and}
  \bibinfo{author}{\bibfnamefont{A.}~\bibnamefont{Arima}},
  \bibinfo{journal}{Phys. Lett. B} \textbf{\bibinfo{volume}{202}},
  \bibinfo{pages}{15 } (\bibinfo{year}{1988}).

\bibitem[{\citenamefont{Horie and Arima}(1955)}]{Horie1955Phys.Rev.778}
\bibinfo{author}{\bibfnamefont{H.}~\bibnamefont{Horie}} \bibnamefont{and}
  \bibinfo{author}{\bibfnamefont{A.}~\bibnamefont{Arima}},
  \bibinfo{journal}{Phys. Rev.} \textbf{\bibinfo{volume}{99}},
  \bibinfo{pages}{778} (\bibinfo{year}{1955}).

\bibitem[{\citenamefont{de~Shalit and Talmi}(1963)}]{Shalit1963}
\bibinfo{author}{\bibfnamefont{A.}~\bibnamefont{de~Shalit}} \bibnamefont{and}
  \bibinfo{author}{\bibfnamefont{I.}~\bibnamefont{Talmi}},
  \emph{\bibinfo{title}{Nuclear Shell Theory}} (\bibinfo{publisher}{Academic
  Press}, \bibinfo{address}{New York}, \bibinfo{year}{1963}).

\bibitem[{\citenamefont{Neyens}(2003)}]{Neyens2003Rep.Prog.Phys.633}
\bibinfo{author}{\bibfnamefont{G.}~\bibnamefont{Neyens}},
  \bibinfo{journal}{Rep. Prog. Phys.} \textbf{\bibinfo{volume}{66}},
  \bibinfo{pages}{633} (\bibinfo{year}{2003}).

\bibitem[{\citenamefont{Yordanov et~al.}(2013)\citenamefont{Yordanov,
  Balabanski, Biero\ifmmode~\acute{n}\else \'{n}\fi{}, Bissell, Blaum,
  Budin\ifmmode \check{c}\else \v{c}\fi{}evi\ifmmode~\acute{c}\else \'{c}\fi{},
  Fritzsche, Fr\"ommgen, Georgiev, Geppert
  et~al.}}]{Yordanov2013Phys.Rev.Lett.192501}
\bibinfo{author}{\bibfnamefont{D.~T.} \bibnamefont{Yordanov}},
  \bibinfo{author}{\bibfnamefont{D.~L.} \bibnamefont{Balabanski}},
  \bibinfo{author}{\bibfnamefont{J.}~\bibnamefont{Biero\ifmmode~\acute{n}\else
  \'{n}\fi{}}}, \bibinfo{author}{\bibfnamefont{M.~L.} \bibnamefont{Bissell}},
  \bibinfo{author}{\bibfnamefont{K.}~\bibnamefont{Blaum}},
  \bibinfo{author}{\bibfnamefont{I.}~\bibnamefont{Budin\ifmmode \check{c}\else
  \v{c}\fi{}evi\ifmmode~\acute{c}\else \'{c}\fi{}}},
  \bibinfo{author}{\bibfnamefont{S.}~\bibnamefont{Fritzsche}},
  \bibinfo{author}{\bibfnamefont{N.}~\bibnamefont{Fr\"ommgen}},
  \bibinfo{author}{\bibfnamefont{G.}~\bibnamefont{Georgiev}},
  \bibinfo{author}{\bibfnamefont{C.}~\bibnamefont{Geppert}},
  \bibnamefont{et~al.}, \bibinfo{journal}{Phys. Rev. Lett.}
  \textbf{\bibinfo{volume}{110}}, \bibinfo{pages}{192501}
  (\bibinfo{year}{2013}).

\bibitem[{\citenamefont{Wood}(2013)}]{Wood2013Physics52}
\bibinfo{author}{\bibfnamefont{J.}~\bibnamefont{Wood}},
  \bibinfo{journal}{Physics} \textbf{\bibinfo{volume}{6}}, \bibinfo{pages}{52}
  (\bibinfo{year}{2013}).

\bibitem[{\citenamefont{Ring}(1996)}]{Ring1996Prog.Part.Nucl.Phys.193}
\bibinfo{author}{\bibfnamefont{P.}~\bibnamefont{Ring}}, \bibinfo{journal}{Prog.
  Part. Nucl. Phys.} \textbf{\bibinfo{volume}{37}}, \bibinfo{pages}{193 }
  (\bibinfo{year}{1996}).

\bibitem[{\citenamefont{Vretenar et~al.}(2005)\citenamefont{Vretenar,
  Afanasjev, Lalazissis, and Ring}}]{Vretenar2005Phys.Rep.101}
\bibinfo{author}{\bibfnamefont{D.}~\bibnamefont{Vretenar}},
  \bibinfo{author}{\bibfnamefont{A.~V.} \bibnamefont{Afanasjev}},
  \bibinfo{author}{\bibfnamefont{G.~A.} \bibnamefont{Lalazissis}},
  \bibnamefont{and} \bibinfo{author}{\bibfnamefont{P.}~\bibnamefont{Ring}},
  \bibinfo{journal}{Phys. Rep.} \textbf{\bibinfo{volume}{409}},
  \bibinfo{pages}{101} (\bibinfo{year}{2005}).

\bibitem[{\citenamefont{Meng et~al.}(2006)\citenamefont{Meng, Toki, Zhou,
  Zhang, Long, and Geng}}]{Meng2006Prog.Part.Nucl.Phys.470}
\bibinfo{author}{\bibfnamefont{J.}~\bibnamefont{Meng}},
  \bibinfo{author}{\bibfnamefont{H.}~\bibnamefont{Toki}},
  \bibinfo{author}{\bibfnamefont{S.}~\bibnamefont{Zhou}},
  \bibinfo{author}{\bibfnamefont{S.}~\bibnamefont{Zhang}},
  \bibinfo{author}{\bibfnamefont{W.}~\bibnamefont{Long}}, \bibnamefont{and}
  \bibinfo{author}{\bibfnamefont{L.}~\bibnamefont{Geng}},
  \bibinfo{journal}{Prog. Part. Nucl. Phys.} \textbf{\bibinfo{volume}{57}},
  \bibinfo{pages}{470} (\bibinfo{year}{2006}).

\bibitem[{\citenamefont{Meng et~al.}(2013)\citenamefont{Meng, Peng, Zhang, and
  Zhao}}]{Meng2013FrontiersofPhysics55}
\bibinfo{author}{\bibfnamefont{J.}~\bibnamefont{Meng}},
  \bibinfo{author}{\bibfnamefont{J.}~\bibnamefont{Peng}},
  \bibinfo{author}{\bibfnamefont{S.-Q.} \bibnamefont{Zhang}}, \bibnamefont{and}
  \bibinfo{author}{\bibfnamefont{P.-W.} \bibnamefont{Zhao}},
  \bibinfo{journal}{Front. Phys.} \textbf{\bibinfo{volume}{8}},
  \bibinfo{pages}{55} (\bibinfo{year}{2013}).

\bibitem[{\citenamefont{Ring}(2012)}]{Ring2012PhysicaScripta14035}
\bibinfo{author}{\bibfnamefont{P.}~\bibnamefont{Ring}}, \bibinfo{journal}{Phys.
  Scr.} \textbf{\bibinfo{volume}{T150}}, \bibinfo{pages}{014035}
  (\bibinfo{year}{2012}).

\bibitem[{\citenamefont{Ring et~al.}(1997)\citenamefont{Ring, Gambhir, and
  Lalazissis}}]{Ring1997Comput.Phys.Commun.77}
\bibinfo{author}{\bibfnamefont{P.}~\bibnamefont{Ring}},
  \bibinfo{author}{\bibfnamefont{Y.~K.} \bibnamefont{Gambhir}},
  \bibnamefont{and} \bibinfo{author}{\bibfnamefont{G.~A.}
  \bibnamefont{Lalazissis}}, \bibinfo{journal}{Comput. Phys. Commun.}
  \textbf{\bibinfo{volume}{105}}, \bibinfo{pages}{77 } (\bibinfo{year}{1997}).

\bibitem[{\citenamefont{Zhao et~al.}(2010)\citenamefont{Zhao, Li, Yao, and
  Meng}}]{Zhao2010Phys.Rev.C54319}
\bibinfo{author}{\bibfnamefont{P.~W.} \bibnamefont{Zhao}},
  \bibinfo{author}{\bibfnamefont{Z.~P.} \bibnamefont{Li}},
  \bibinfo{author}{\bibfnamefont{J.~M.} \bibnamefont{Yao}}, \bibnamefont{and}
  \bibinfo{author}{\bibfnamefont{J.}~\bibnamefont{Meng}},
  \bibinfo{journal}{Phys. Rev. C} \textbf{\bibinfo{volume}{82}},
  \bibinfo{pages}{054319} (\bibinfo{year}{2010}).

\end{thebibliography}
\end{document}